\PassOptionsToPackage{unicode}{hyperref}
\PassOptionsToPackage{hyphens}{url}
\PassOptionsToPackage{dvipsnames,svgnames,x11names}{xcolor}
\documentclass[%
french,english%
,nocjs%
]{cjs-rcs-article}

\usepackage{longtable,booktabs,array}
\usepackage{calc} 
\usepackage{etoolbox}
\makeatletter
\patchcmd\longtable{\par}{\if@noskipsec\mbox{}\fi\par}{}{}
\makeatother
\IfFileExists{footnotehyper.sty}{\usepackage{footnotehyper}}{\usepackage{footnote}}
\makesavenoteenv{longtable}
\usepackage{graphicx}
\makeatletter
\newsavebox\pandoc@box
\newcommand*\pandocbounded[1]{
  \sbox\pandoc@box{#1}%
  \Gscale@div\@tempa{\textheight}{\dimexpr\ht\pandoc@box+\dp\pandoc@box\relax}%
  \Gscale@div\@tempb{\linewidth}{\wd\pandoc@box}%
  \ifdim\@tempb\p@<\@tempa\p@\let\@tempa\@tempb\fi
  \ifdim\@tempa\p@<\p@\scalebox{\@tempa}{\usebox\pandoc@box}%
  \else\usebox{\pandoc@box}%
  \fi%
}
\def\fps@figure{htbp}
\makeatother
\providecommand{\tightlist}{%
  \setlength{\itemsep}{0pt}\setlength{\parskip}{0pt}}
\makeatletter
\@ifpackageloaded{caption}{}{\usepackage{caption}}
\AtBeginDocument{%
\ifdefined\contentsname
  \renewcommand*\contentsname{Table of contents}
\else
  \newcommand\contentsname{Table of contents}
\fi
\ifdefined\listfigurename
  \renewcommand*\listfigurename{List of Figures}
\else
  \newcommand\listfigurename{List of Figures}
\fi
\ifdefined\listtablename
  \renewcommand*\listtablename{List of Tables}
\else
  \newcommand\listtablename{List of Tables}
\fi
\ifdefined\figurename
  \renewcommand*\figurename{Figure}
\else
  \newcommand\figurename{Figure}
\fi
\ifdefined\tablename
  \renewcommand*\tablename{Table}
\else
  \newcommand\tablename{Table}
\fi
}
\@ifpackageloaded{float}{}{\usepackage{float}}
\floatstyle{ruled}
\@ifundefined{c@chapter}{\newfloat{codelisting}{h}{lop}}{\newfloat{codelisting}{h}{lop}[chapter]}
\floatname{codelisting}{Listing}

\makeatother
\makeatletter
\@ifpackageloaded{caption}{}{\usepackage{caption}}
\@ifpackageloaded{subcaption}{}{\usepackage{subcaption}}
\makeatother

\title{Guide to Teaching with Data in Statistics and Data Science
Classrooms}

\author[orcid=0000-0002-8007-934X,email=mdogucu@fas.harvard.edu,corresponding]{Mine \surname{Doğucu}}
\affil{Department of Statistics, Harvard University, Cambridge,
MA, United States of America}
\author[orcid=0000-0003-2847-8180,]{Catalina \surname{Medina}}
\affil{Department of Mathematics and Data Science, California State
University Channel Islands, Ventura, CA, United States of America}
\author[orcid=0009-0003-9912-5064,]{Crystal \surname{Lewis}}
\affil{Crystal Lewis Consulting LLC, St.~Louis, MO, United States of
America}

\begin{englishabstract}
Statistics and data science educators have been concerned about data use
or lack thereof in the classroom for decades. For many educators, the
question is not whether to use data or not. They often want to know
whether they should give clean or messy data, and whether they should
spend the time to find a real dataset or simulate it. If they decide to
find real data, then other questions follow about where to find it and
how to provide it to students. If they provide data, then the question
remains whether students will find the data engaging or not or if it
will have the properties necessary to demonstrate class concepts. We
provide empirically and theoretically supported recommendations on data
use. Classroom data are described in different ways (e.g., real, raw,
made-up, artificial, fictitious, synthetic, simulated, toy, and messy).
We provide definitions of different types, sources, formats, and shapes
of data and provide distinctions between classifications. We share
resources for educators in bringing data to their classrooms.
\end{englishabstract}

\begin{keywords}
\item Data science education
\item Statistics education
\item Use of real data
\item Data simulation
\item Student engagement
\end{keywords}

\begin{document}
\maketitle
\suppressfloats[t] 

\section{Introduction}\label{sec-intro}

Statistics and data science educators have been concerned about data use
or lack thereof in the classroom for decades. At around the turn of the
century, Higgins argued for undergraduate statistics programs to focus
on nonmathematical aspects of the discipline and have students work with
data to prepare them for a career as, what he calls, a data specialist
\citeyearpar{higgins1999nonmathematical}. Use of data is also an
important part of data literacy to pave the way for active and informed
citizenship \citep{weiland2026conceptualizing}.

In addition to whether to use data or not, another question that
statistics educators had to deal with is whether to use real data or
not. For instance in his book \emph{An Introduction to Statistical
Methods in the Behavioral Sciences} Elzey wrote ``All data presented in
this book are fictitious'' \citeyearpar{elzey1976introduction}. The use
of ``fictitious'' data back in the 1970s was harshly critiqued by
\citet{meyer1978statistics} precisely with the question ``How can we
teach methods so divorced from substantive import that we do not even
have to present one set of real data?'' Considering that the book is
focused on Statistical Methods in the Behavioral Sciences, he goes on to
criticize the fact that the fictitious data is labeled as ``anxiety
scores'', ``test scores'', etc., and that if the author were to change
the labels he could entitle the book as ``Statistical Methods in \_\_\_
(Fill in the blank).'' Data context along with the data being real or
not has been a question of concern as well.

Even when educators use data that originated as real data, they might
modify the data to demonstrate specific statistical content. For
instance referring to the accompanying textbook R package \{bayesrules\}
\citep{dogucu2021bayesrules}, Johnson et al. \citep{johnson2022bayes}
state that they selected the data from larger resources using non-random
methods to match their pedagogical goals. They warn the readers not to
use the data for research purposes and refer them to the original and
complete data source should they be interested in conducting research.
This raises an important distinction between use of data for research
and for teaching.

Over the recent years, especially with an emphasis on reproducibility
and research conduct, there have been several important guides
\citep[e.g.,][]{borghi2018support, borer2009some} and books
\citep[e.g.,][]{corti2020managing, lewis2024data} in handling data for
research purposes. In this manuscript we borrow from this literature to
build the guidelines for data use in the classroom. Our first aim in
writing this manuscript is providing empirically and theoretically
supported recommendations for the classroom on data use. When empirical
and theoretical evidence is lacking we provide positional statements
based on our teaching experiences.

In our combined years of teaching statistics, working with data, and our
reading of the literature we have encountered several adjectives to
describe data including but not limited to real
\citep{gould2010statistics}, raw \citep{singer1990improving}, made-up
\citep{medina2026engaging}, artificial \citep{dunn2024impact},
fictitious \citep{fernandez2024technology}, synthetic
\citep{bulmer2022role}, simulated \citep{blejec2002teaching}, toy
\citep{timbers2026ten}, and messy \citep{messy} specifically for data
used in the classroom. Our second aim in writing this manuscript is
providing clear definitions of different types, sources, formats, and
shapes of data for the statistics and data science classrooms and
providing distinctions between data classifications. As we provide
different definitions for each type of data, we discuss challenges and
opportunities associated with each data type and provide guidelines on
the types of learning contexts in which different data might be
beneficial.

For many statistics and data science educators, the question is not
whether to use data or not. They often want to know whether they should
give clean or messy data, and whether they should spend the time to find
a real dataset or they should simply simulate it. If they decide to find
real data, then several other questions follow about where to find the
data and how to provide it to students. If they provide data to their
students, then the question remains whether students will find the data
engaging or not or if it will have the properties necessary to
demonstrate class concepts. Our third aim in writing this manuscript is
providing resources for statistics and data science educators in
bringing data to their own classrooms. This manuscript will guide
statistics and data science educators in answering several questions
related to data use in the classroom including but not limited to 1.
\emph{Why} instructors should use data in the classroom; 2.\emph{What}
kind of data they should use for different teaching contexts. 3.
\emph{How} they can find, store, and distribute data for their students.

To address these questions we start with definitions in
Section~\ref{sec-definitions} of different data types and discuss
challenges and opportunities related to each. Later in
Section~\ref{sec-recommendations} we provide recommendations for
classroom practice related to data use. The manuscript is written with
undergraduate educators in mind but many statements will generalize to
teaching of statistics in K-12 (especially in grades 8-12) graduate
education, and training of practitioners.

\section{Definitions, Challenges, and
Opportunities}\label{sec-definitions}

\subsection{Real Data}\label{real-data}

For decades, there have been many statistics educators calling for use
of real data when teaching
\citep{gehrke2021statistics, neumann2013using, cobb1993reconsidering, singer1990improving, singer1988opening}.
The Guidelines for Assessment and Instruction in Statistics Education
(GAISE) College Report also recommends instructors to ``integrate real
data with a context and purpose'' \citeyearpar{gaise2016}. Real data
always comes with its context. We will delve deeper into the context of
data and its implications for the classroom in
Section~\ref{sec-context}.

\begin{figure}[tbp]

\centering{

\includegraphics[width=0.85\linewidth,height=\textheight,keepaspectratio,alt={Flow diagram: raw data is cleaned into one cleaned dataset, which is then split into three teaching datasets.}]{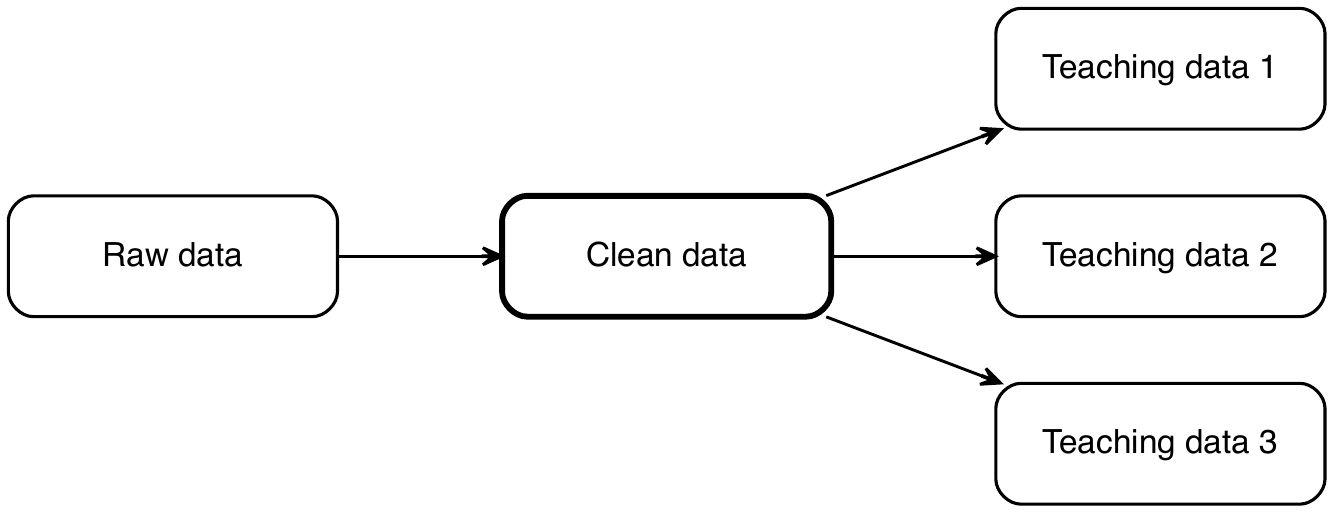}

}

\caption{\label{fig-real-data}Classification of real data for teaching
purposes}

\end{figure}

Without a doubt real data should be embraced by statistics educators in
the classroom, especially for courses that are less theoretical and
focused more on data analysis. Real data can make an appearance in the
classroom in different ways. In a research setting Lewis
\citeyearpar{lewis2024data} classifies data that researchers work with
as raw data, clean data, and analysis data. With an inspiration from
this classification we classify real data as raw data, clean data, and
teaching data for teaching purposes as shown in
Figure~\ref{fig-real-data}.

Raw data is often the form of data that researchers collect and is
stored as it is collected. It can potentially include identifiers,
missing data, duplicate data, or data entry errors (e.g., text values
entered into fields intended to be numeric). Raw data is often not
published. For instance, statistical agencies do not publish raw data
\citep{fellegi1988}. The cleaned data may directly be provided by
researchers to remove identifiers or data errors. The instructors might
also clean data from a raw source to make the messy data more human
readable by changing variable names, replacing numerical categories with
their labels or recoding 999, -777 or similar values as missing.

Instructors then might make many different modifications to cleaned data
to make the teaching and learning process easier arriving at what we
call teaching data. For instance one common modification includes
subsetting data. The subsetting might include practices such as
selecting only a few variables that students can focus on, taking a
random sample of observations to overcome size restrictions of learning
or data science platforms (e.g., GitHub data storage limit) or taking a
nonrandom sample (e.g., the aforementioned example from \emph{Bayes
Rules!} book) to demonstrate a specific statistical content. Depending
on different teaching contexts, an instructor might have different sets
of teaching data derived from a single cleaned data source.

In using real data the goal can be for students to encounter real data
cleaning problems (e.g., mismatch of categories across different
variables) and real data modeling problems (e.g., portraying the
difficulty of modeling). If the learning goal is the former then
providing raw data is necessary and if it is the latter then cleaned
data can be provided.

It is logistically impossible to have students work with raw or cleaned
data on a regular basis. To get to the specific teaching goals, it will
make sense to generate teaching data. If every class started with data
extraction/collection, and then data cleaning just to expose students to
the ``real'' experience of being a data scientist, then it is possible
that any specific statistical content would not be covered within
limited class time. Thus it is important for educators to embrace all
forms of real data throughout different teaching contexts.

There is also a time and place, for exposing students to the full data
cycle from real data cleaning problems to real data modeling problems
simultaneously. The full data cycle experience requires a longer time
period than a class or homework assignment.

Project-based learning in statistics
\citep{macgillivray2011teaching, martonosi2016survey} and data science
classrooms \citep{ccetinkaya20225ws, lasser2021introductory} is a great
way to expose students to the full data cycle. In addition, external
activities such as DataFest \citep{noll2023insights, gould2014datafest}
and datathons \citep{anslow2016datathons} can also provide opportunities
for students to practice the full data cycle.

\subsection{Simulated Data}\label{simulated-data}

Simulated data has also made an appearance with different names in the
literature including artificial \citep{singer1988opening}, fictitious
\citep{weiland2024culturally}, and synthetic
\citep{pfannkuch2018innovations}. The two common formats of simulated
data we encounter in the classroom are simulated data that is
empirically informed and simulated data that is made up
\citep{medina2026engaging}. The former uses a data context and prior
research to simulate data based on a reasonable model with some random
mechanism. The latter, however, has no concern about the numbers
representing reasonable values. In fact in the latter case, one can
encounter variables with generic names such as \(X_1\), \(X_2\), and so
on.

Simulated data can be generated by instructors or students and can serve
many purposes in the classroom. First and foremost, students can
simulate data in learning scenarios where a practitioner would also
simulate data, in topics such as posterior approximation in Bayesian
analysis or power calculation in frequentist analysis. Computer
simulations can also be useful in advanced courses with proofs.
Simulations can solidify concepts in the theoretical proofs
\citep{blejec2002teaching}. The advances in computing power also allow
for more room in the curriculum for simulations.

Many statistics classroom activities involve data generated with playing
cards \citep{eckert1994teaching}, dice \citep{koparan2016using}, M\&M's
\citep{schwartz2013teaching} or computer code and applications
\citep{potter2016web}. In their comparison of students who learn
simulation with hands-on tools followed by computers and those who learn
simulation by computers alone, Hancock and Rummerfield see better
improvement from the first midterm to the final exam in the group that
learned simulation first with hands-on objects and thus recommend that
instructors start teaching simulation based topics first with physical
objects \citep{hancock2020simulation}. In recent decades,
simulation-based inference has become an important part of teaching
statistics, especially in introductory statistics courses
\citep{fernandes2026investigating, chance2016student, rossman2014using}
thus the use of data generated by physical tools and computers has
become an even more important part of the statistics classrooms.

Simulated data can bring a wealth of discussion on randomness to the
classroom. A great example of this is by Deborah Nolan whose classroom
activity appeared on the Radiolab podcast
\citep{radiolab2009stochasticity}. She makes half her students generate
100 random coin flips and half her students make up 100 coin flips while
she stands outside the classroom. She then returns to the classroom to
guess which sequence was randomly generated. She shares her secret to
guessing correctly is that when people make up data they do not write
long runs of heads or tails. \citet{schneiter2023leveraging} explore how
large class sizes can be leveraged to facilitate student-driven
simulation where the class can be the population of interest and samples
can be collected from students in the class to explore inference,
effects of sample size, experimental design, and messy data.

Instructors sometimes have to make up data because the nature of the
context is fictional itself. These fictional scenarios might appeal to
students' interest in fiction or fantasy. One such example is the set of
dragonistics data cards \citep{robertson2023building} that feature a
unique dragon on each card with their features such as strength level,
age, height etc. Students then can use these cards to sort them, group
them, etc.

Another excellent use of fictionally simulated data is the game
\emph{Islands} which provides a virtual environment of three islands
with residents whom students can collect data from
\citep{bulmer2011life}. Although the data is not real, there are many
aspects of the virtual environment that are ``real-like'' that allow
students to learn statistical skills including asking for consent for
data collection, waiting for data to be collected, and the complexity of
sampling decisions that goes into data collection. The Islands game can
be used in project-based assessments \citep{tong2024bridging}.

Although overall teaching recommendations include real data use in the
classroom which we agree with, in this section we have also seen several
uses of simulated data examples that can be appropriate and effective in
teaching statistics and data science. The use of simulated data when one
would methodologically simulate data is nonnegotiable in the classroom.
However, for applied examples in the classroom, instructors can make a
choice between real data and simulated data. When real data for a
specific learning context is not available for different reasons (e.g.,
COVID-19 lockdowns \citep{dunn2024impact}) then instructors may choose
to use simulated data such as the Islands game. In these cases where
data is simulated for instructional purposes, Bulmer and Coote
\citeyearpar{bulmer2022role} recommend three dimensions to be satisfied:
relevance, realism, and richness.

\subsection{Data Context and Relevance}\label{sec-context}

As Cobb and Moore said ``Data are not just numbers, they are numbers
with a context'' \citep{cobb1997}.

Real data arise from particular circumstances, and understanding those
circumstances helps students interpret what the data can tell us. When
teaching with real data, teachers should therefore make intentional
decisions about what contexts they select and how they share the context
with their students. Research has identified benefits associated with
students finding meaning and interest in the data they encounter in
their course. \citet{neumann2013using} found students associated
real-life data with greater relevance, interest, learning, memory,
motivation, engagement, and understanding. \citet{medina2026engaging}
identified further themes related to students interacting with contexts
they were interested in such as better experience working on assignments
(e.g., more fun, easier, motivating) and desire to learn more about the
data context.

Researchers have used their own experiences and/or research to provide
recommendations on ways instructors can identify real data with contexts
that students find interesting or relevant. Domain-specific courses like
biostatistics can draw from discipline-specific peer-reviewed literature
\citep{brearley2018tshs}. Other ways to identify relevant contexts can
be based on the students' location \citep{brooks2021towards} or cultural
relevance by drawing upon students' lived experiences with culturally
relevant data \citep{weiland2024culturally}. Researchers have suggested
various additional ways data may be relevant to students like through
coffee sales \citep{sole2017s}, everyday technology
\citep{gould2010statistics}, and popular culture
\citep{gunderman2020developing}. \citet{medina2026engaging} provide
further ways students identified data as interesting to them, and found
providing students choice of context can be one avenue to helping
students connect with their assignment more.

It warrants mentioning that there has been work that found a potential
for an increased burden when introducing real data contexts.
\citet{phadke2024} found a slight decrease in average assessment scores
in an experiment comparing an old version and a modified version of the
assessment that differed by context. In the modified version they
incorporated more real data with contexts, but there are other
differences between the assessments as well. Since there are multiple
features differing between the assessment versions, further research is
required to determine which features contributed to the difference in
performance. Pfannkuch's \citeyearpar{pfannkuch2011} case study findings
suggest data context can support interpretation, but it can also divert
attention during the construction of concepts. A caution when
introducing a new data context is considering the cognitive load
required for students to understand the context. Unfamiliar topics may
warrant guided discussion before independent work. Another consideration
to alleviate possible increased burden from comprehending new contexts
could be to introduce new contexts intentionally and maybe revisit
previously introduced contexts when students are learning an especially
difficult new statistical concept.

\subsection{Data Collection}\label{sec-collection}

Data collection can occur in multiple ways in the statistics and data
science classrooms. One of the ways is that the instructor can collect
data from their students which can provide real data for classroom use
\citep{stork2003teaching}. Since the data are about the students, they
may possibly find it relevant and interesting to work with. An example
is a survey developed by \citet{neumann2010development} which has
questions about the students such as their height, weight, and birth
date. It also asks them to guess the instructor's age and number of
children. They consider different types of variables (numeric,
categorical) to be an important feature of the survey. It also allows
for students to get to know the instructor.

Another example is by \citet{conlon2026enhancing} who have replaced
traditional assessments with projects that use data from a student
survey on plastic recycling behaviors. Needless to say, instructors who
plan on collecting student data via surveys should consider what is
appropriate to ask students, what students may feel comfortable sharing
on a survey, and how anonymity will be preserved.

Another way data can be collected in the classroom is by the students
themselves in which case data collection itself would be one of the
learning goals of the class. Data collection led by students can promote
active learning \citep{strangfeld2013promoting}. In their comparison of
a regression activity that has some students interact with a web app and
collect their own data and some students work with pre-collected data on
a paper worksheet, \citet{kuiper2025greenhouse} conclude that the
students in the first group show higher gains in knowledge.

An excellent example of students collecting data is by
\citet{woodard2023five} which lays out five hands-on experiments for a
design of experiments course. Her examples include questions such as
``Are dominant arms stronger than non-dominant arms?'' and ``Do people
remember better if they hear a list, see a list or both?'' for which
students need to think about the design of the experiment, plan the data
collection process, collect data and run into problems while collecting
data.

Student data can also appear in the classroom through the activities
that students conduct. For instance, \citet{kuiper2015using} suggest use
of computer game based simulations in teaching of inference. Stat2Games
(https://www.stat2games.sites.grinnell.edu/) is an online collection of
games that are designed for statistics classrooms. In a way the game
data is similar to the aforementioned fictitious Islands game data. The
games consist of fictional scenarios such as crop-growth and car racing.
\citet{ricci2026see} use the car racing game in teaching of the Bayesian
Beta-Binomial model for instance. What is worth noting is that even
though the games consist of fictitious scenarios (e.g., car racing), the
data generated is real (e.g., students' completion time of the online
car race). This data is generated by students themselves.

\subsection{Data Scale (Big Data vs.~Small
Data)}\label{data-scale-big-data-vs.-small-data}

We are generating more data every day than we did yesterday. The
emergence of big data has brought one of the biggest changes for the
statistics classrooms. Computational skills and toolkits for working
with big data have become extremely important for the preparation of
students for the data science workforce \citep{franccois2020big}. There
have been many institutions offering courses specifically focused on big
data analytics (e.g., \citep{asamoah2017preparing, eckroth2018course}).

Some specific changes that statistics educators have considered or made
in the last two decades include teaching of databases and SQL
\citep{stander2017enthusing}, parallel computing
\citep{nolan2010computing}, and distributed computing
\citep{alexander2026reflections}.

Emphasizing the importance of teaching big data skills in the age of
data revolution, Ridgway \citeyearpar{ridgway2016implications} calls for
educators to place less emphasis on small samples and considers unstable
estimates and generalization to be key problems with working with small
sample data. While we do generate more and more data every day, the
issues with small sample data are still important for students to know,
so it is important for them to encounter both big data and small data in
their educational journey.

Encountering small data still remains important for several reasons.
First, some research is small sample by its nature, such as rare disease
research \citep{griggs2009clinical}. Thus it is important for students
to see the value of research conducted both with big data and small
data, seeing scientific contributions of both. Second, statisticians can
be trained in both quantitative research and qualitative research
methods as well as mixed-methods research. For instance qualitative
research can be extremely helpful in survey design
\citep{mauceri2014mixed}. By their nature, qualitative research studies
employ small sample sizes. Since statistics students spend more time
learning quantitative methods, it is of utmost importance for them to be
aware of qualitative research not only for their own data skills but
also to develop trust in science utilizing different methods than their
own discipline.

Given the limited time in the curriculum perhaps providing an extensive
training in qualitative research in undergraduate statistics and data
science programs may not be feasible. However, it still makes sense for
students to become aware of small sample size research scenarios briefly
when possible.

\subsection{Data Curation and
Dissemination}\label{data-curation-and-dissemination}

In situations where the instructors create datasets for their students
via organizing, cleaning, and maintaining, instructors start wearing a
new hat: data curator. Every statistics educator does not necessarily
need to become a data curator. Data curation can require a lot of time
and involves multiple steps \citep{perry2022measuring}. Even going
through the simplified steps shown in Figure~\ref{fig-real-data}, data
curators need to make several decisions. Some educators can choose to
rely on pre-prepared data as is for their teaching.

The first question in curating data is \emph{where} to find data. Many
government agencies and international organizations, such as Fisheries
and Oceans Canada and the United Nations, provide statistical data
online.\\
Data curator educators with specific courses such as econometrics, would
thus need to familiarize themselves with agencies that provide economics
data such as the Bureau of Labor Statistics in the United States, the
Organisation for Economic Co-operation and Development, and so on.

There are also many generalist dataset repositories online that
instructors can turn to including Kaggle \citep{kaggle}, Harvard
Dataverse \citep{harvarddataverse}, UCI Machine Learning Repository
\citep{uci}, Inter-university Consortium for Political and Social
Research (ICPSR) \citep{icpsr}, Figshare \citep{figshare}, and Zenodo
\citep{zenodo}, as well as domain specific repositories such as LDbase
\citep{hart2020ldbase} which caters specifically to education and
developmental science data.

Many of the repositories have created resources to help users navigate
their sites. For instance, if teaching with ICPSR, readers may want to
watch the webinar by \citet{hoelter}. While using data from any data
repository, educators should be especially careful with paying attention
to data source to ensure that the data are not made-up. Kaggle in
particular hosts several datasets with unclear data source citations.
Data curators should either have a bookmarks folder or some other way of
keeping a list of data sources. Although these lists help, maintaining
them can be arduous as sources disappear or links change.

With search capabilities increasing through tools that use large
language models, instructors can also possibly turn to these tools when
searching for novel datasets. These tools are more likely to give more
commonly used datasets such as \texttt{penguins}, \texttt{mtcars},
\texttt{iris}, and \texttt{flights}. Thus they need to be specifically
instructed on dataset context, size, format, etc.

Data curators will need to pay attention to \emph{what license} the data
were shared under. Although this is not legal advice, use of data for
educational purposes may be more permissive than use for commercial
purposes. Thus whether educators curate data to use with their students
in the class vs.~to share data with their textbook might have different
permissions.

A data curation process for teaching and learning should include: a
close study of data context as it can make a difference in statistical
analysis decisions \citep{finch2023}; well thought out variable names
and file names \citep{finch2023}; and a well documented README file with
citations of data source and host \citep{dogucu2022tools}.

Once the data curators are ready to disseminate the data then they need
to decide \emph{how} to disseminate it. There are many options including
but not limited to sharing the data with students on learning management
systems (e.g., Canvas), cloud links (e.g., Google Drive), R packages,
and GitHub repositories.

Since local server file paths, GitHub repositories, and Cloud storage
files are not permanent and can become inaccessible over time, if
possible, teachers should provide data in downloadable packages from
package repositories. For example, R packages, available from CRAN,
provide an open-access and reproducible way to disseminate data for
teaching purposes \citep{dogucu2025reproducibility}. Statisticians
already have several datasets in the R ecosystem. For instance there is
the \{datasets\} package with several datasets in base R \citep{R-base}.
In fact the contexts of these datasets have been studied by
\citet{finch2023}. However, in their analysis, they state that the
contexts of the datasets are impoverished and at times wrong. Relying on
general purpose R datasets can potentially be misleading and instead
relying on datasets that are specifically curated for teaching and
learning might be a better alternative. R packages that supplement
textbooks can be such examples (e.g., \{openintro\} \citep{openintro},
and \{ISLR\} \citep{ISLR}).

Although data curation is time-consuming and challenging it comes with
several benefits. An instructor who can curate data can possibly curate
data that will be interesting for and relevant to their own students.
Data curation is also a skill for students to know. The better an
instructor gets at this, the better they will be able to train their own
students in data curation.

Last but not least, when data curation work is shared publicly, the
impact of the work on statistics education can scale to a much larger
population beyond instructors' own students. For instance, the
\{palmerpenguins\} package in R provides ecological data on penguins
\citep{horst2022palmer} and has been downloaded 4,056,976 times from
CRAN since its first appearance on 2020-07-23 until today (2026-10-30).
We cannot speculate how many of the downloads were due to ecological
interests in penguins but considering its wide use in teaching
demonstrations, it would not be wrong to conclude that good teaching
demonstration datasets such as \{palmerpenguins\} are impactful and can
reach many educators and learners.

\subsection{Data Ethics}\label{sec-ethics}

Bringing data to the classroom requires bringing ethical considerations
along with the data itself.

While data science education as an interdisciplinary field may
disproportionately focus on statistical, computational, and technical
skills, ethics is considered to be an important pillar of data science
\citep{bednarowska2026bringing}. Ethical issues can be and should be
covered in multiple stages of the curriculum. From the data collection
process (e.g., consent \citep{raman2023integrating}), to modeling (e.g.,
using inappropriate Bayesian priors \citep{utts2021enhancing}), and to
algorithms (e.g., algorithmic bias \citep{baumer2022integrating}),
ethical issues appear throughout the data cycle and hence should appear
throughout the curriculum to reflect the need.

Statistics educators have several ways, angles, and examples of teaching
ethics. One is through discussion of case studies as seen in
\citet{tractenberg2023ethical}. Ethics can be taught on a wide spectrum
from philosophical foundations to technical applications. Readers can
find examples from a wide spectrum of data science ethics courses in
\citet{colando2024philosophy}. Considering the historical ties of some
statisticians with the eugenics movement, instructors can also consider
connecting historical mistakes with current ethical issues
\citep{kennedy2024teaching}. Last but not least,
\citet{raman2023integrating} have extensive guidelines on incorporating
ethics into introductory statistics courses with several prompts for
students to consider.

\section{Recommendations}\label{sec-recommendations}

In this section we provide recommendations on data use in statistics and
data science classrooms. The students enrolled in statistics and data
science courses can be training as statisticians, data scientists or
generalists in their degrees, sometimes training as part of a specific
discipline (e.g., Introduction to Statistics for Computer Science).
Given the variety of levels and audiences of the courses taught by
statistics educators, we do not envision a single class where all of
these recommendations can be met. However, these recommendations will be
relevant in a wide range of classes and in degree programs.

\begin{enumerate}
\def\labelenumi{\arabic{enumi}.}
\tightlist
\item
  \textbf{Students must use data from different sources}
\end{enumerate}

Students should work with data from different sources including but not
limited to federal statistics agencies, research studies, and industry
data releases. Students can collect data from its source or use
secondary data. This can expose them to additional topics such as ethics
of data collection and understanding data codebooks for secondary data.
The data can come from a static source such as csv files or a dynamic
source \citep{hardin2016dynamic} through web scraping
\citep{dogucu2021web} or API access \citep{fergusson2021traversing}.

\begin{enumerate}
\def\labelenumi{\arabic{enumi}.}
\setcounter{enumi}{1}
\tightlist
\item
  \textbf{Students must work with different types of data: numeric,
  categorical, text, image, and sound}
\end{enumerate}

Throughout their statistics training, students should get to work with
different types of data including numeric and categorical variables. In
addition, even at the introductory level students can be exposed to
working with text data (i.e., strings). Image data has also become an
important part of statistics. For instance, in 2024 a new journal called
\emph{Statistics and Data Science in Imaging} was founded
\citep{vannucci2024statistics} bringing together statistical methods for
the analysis of imaging data across various disciplines (e.g., medical,
forensic, and satellite imaging). This shows the expansion in importance
and relevance of statistics in image analysis. Although this specific
journal may not speak to the level of undergraduate students, there are
learning activities that can be adopted at this level \citep[e.g.,
breast cancer image classification in][]{timbers2022data} and even at
younger levels \citep{fergusson2024using}.

\begin{enumerate}
\def\labelenumi{\arabic{enumi}.}
\setcounter{enumi}{2}
\tightlist
\item
  \textbf{Students must work with different sizes of data}
\end{enumerate}

With increases in computational storage, availability of tools and
computing power in handling big data, teaching of big data has become
more common at the undergraduate level. These changes allow for and
requires statistics and data science students to be exposed to topics
(e.g., distributed and parallel computing) and tools (e.g., Hadoop) that
fall outside of traditional statistics. Since many statisticians are not
typically trained in handling of large data, departments will either
need to rely on computer science and similar departments to teach
students these topics, or they will have to provide opportunities for
their instructors to receive training in computational topics. With the
emergence of data science, instructors with statistics backgrounds have
relied on pre-conference workshops and professional development
opportunities to acquire computational data skills
\citep{demirci2025understanding}. Students must also be aware of
scientific contributions of studies with small sample size by design
(e.g., rare disease research and qualitative research).

\begin{enumerate}
\def\labelenumi{\arabic{enumi}.}
\setcounter{enumi}{3}
\tightlist
\item
  \textbf{Students must work with different formats of data}
\end{enumerate}

Despite the dominance of R software \citep{R-base} among statisticians,
there are many different software that are used by statisticians,
statistical agencies, and organizations including but not limited to
Python \citep{python}, SAS \citep{SAS}, Stata \citep{Stata}, and SPSS
\citep{SPSS}. In addition to working with typical data formats such as
\texttt{.csv}, students should be provided opportunities to work with
different data formats that are commonly and less commonly found in the
data world such as \texttt{.sav}, \texttt{.sas7bdat}, etc. As students
are introduced to big data, they will also encounter different data
formats (e.g., parquet).

\begin{enumerate}
\def\labelenumi{\arabic{enumi}.}
\setcounter{enumi}{4}
\tightlist
\item
  \textbf{Students must work with different data contexts}
\end{enumerate}

It is important for students to see relevance of statistics in different
disciplines. One way educators can achieve this is through using data
from different contexts throughout their courses. One exception to this
might be statistics courses that are specific to a discipline (e.g.,
Introduction to Statistics for Economics, Introduction to
Biostatistics). Even in these classes instructors can show the relevance
of statistics within subdisciplines (e.g., infectious disease modeling,
clinical trials) of the specific disciplines.

\begin{enumerate}
\def\labelenumi{\arabic{enumi}.}
\setcounter{enumi}{5}
\tightlist
\item
  \textbf{Instructors must adhere to reproducibility practices when
  providing data to students}
\end{enumerate}

Similar to reproducible research practices, reproducible teaching
practices are important \citep{dogucu2022tools}. Metadata (e.g., data
origin, codebooks, changelogs) is especially important in maintaining
data sources from year to year. Having reproducible practices is also
important in updating data from year to year
\citep[e.g.,][]{amaliah2022journey}. When possible, providing data in R
packages and submitting to CRAN might eliminate the problems with
permanence of file location. Last but not least, generation of simulated
data should utilize a random seed for reproducibility of data.

\begin{enumerate}
\def\labelenumi{\arabic{enumi}.}
\setcounter{enumi}{6}
\tightlist
\item
  \textbf{Higher education promotion and review committees should
  consider data contributions as part of promotion portfolios}
\end{enumerate}

Throughout this manuscript, we have shared many benefits of data use in
the statistics and data science classroom and many examples from the
statistics education community. These examples and many more data
sources exist thanks to an educator and/or a researcher who have put the
time and energy into disseminating the data. Many before us have
reiterated the importance of recognizing data contributions in hiring,
promotion, and tenure \citep{puebla2024ten, mayernik2016research} from
research perspectives. We believe this is as important for teaching
perspectives as well.

\section{Closing Remarks}\label{closing-remarks}

Teaching statistics is hard. Many perceive statistics as hard and boring
\citep{bromage2022teaching}. Teaching statistics with data requires
extensive preparation time and thought put in by the instructor
\citep{willett1992providing}. Data with real applications in the
classroom can possibly make statistics fun, engaging, palatable, and
interesting
\citep{willett1992providing, meng2009desired, pavlenko2024teaching}.

It is impossible for a learner's every single data encounter to expose
them to the full data cycle, including data collection/extraction,
wrangling, and analysis. A student studying for a degree in statistics
or data science should encounter different data types, formats, and
contexts in different learning contexts. Instructors of statistics and
data science need to keep an open mind on the purpose data serves in
each class, assessment, or external activity and make decisions on
\emph{why}, \emph{how}, and \emph{when} specific data is used. Thus
rather than having clear cut rules on what data can or cannot be used,
we provided this comprehensive guide for instructors to consider the
challenges and opportunities associated with different data scenarios.
When possible, we provided, examples, resources, and recommendations to
overcome the challenges to make students' data experiences richer while
easing the burden of the instructors.

\section{AI Statement}\label{ai-statement}

The authors have outlined and have written the first draft of the
manuscript. Each then took turns to edit it for clarity, grammar, and
correctness. Generative AI tools (Opus 5.5) were used to comply with the
LaTeX template of the journal, search for some specific publications and
their BibTeX citations, write R code generating the diagram in
Figure~\ref{fig-real-data}, and to do final editing for clarity and
grammar.

\makebackmatter

\bibliography{bibliography.bib}

\end{document}